\documentclass[%
 preprint,
 amsmath,amssymb,
 aps,
]{revtex4-2}
\usepackage[mathlines]{lineno}
\usepackage{graphicx}
\usepackage{dcolumn}
\usepackage{bm}
\usepackage{systeme}

\usepackage[font={small}]{caption}
\usepackage[utf8]{inputenc}
\usepackage{natbib}

\begin{document}

\title{Reconstruction of event kinematics in semi-inclusive deep-inelastic scattering using the hadronic final state and machine learning}
\author{Connor Pecar}
\affiliation{%
 Duke University
}%
\author{Anselm Vossen}
\affiliation{%
 Duke University
}%

\author{Presented at DIS2022: XXIX International Workshop on Deep-Inelastic Scattering and Related Subjects, Santiago de Compostela, Spain, May 2-6 2022}
\maketitle
\section{Introduction}
Deep-inelastic scattering (DIS) of electrons off ions is at the forefront of experimental efforts to probe the internal structure of nucleons and nuclei and will be a primary focus of study at the Electron-Ion Collider. In semi-inclusive DIS, selected particles produced by the fragmentation of the struck quark are observed in coincidence with the scattered electron, \(e(k) + N(P) \rightarrow  e(k') + h(p) + X\), resulting in observables which provide access to a convolution of parton distribution functions (PDFs), describing the momentum of partons within the nucleon, and fragmentation functions (FFs), describing the probability of producing a final state particle with some momentum from the struck quark in the factorization approach. \cite{bacchetta_dalesio_diehl_miller_2004}

\begin{figure}[h]
    \centering
    \includegraphics[width=8cm]{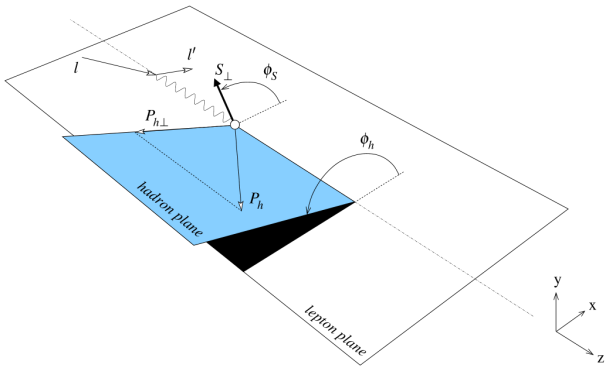}
    \caption{Definition of semi-inclusive kinematic variables in target rest frame, reprinted from \cite{bacchetta_dalesio_diehl_miller_2004}. }
    \label{fig:sidis}
\end{figure}

The kinematic variables describing the DIS process, with center of mass energy squared \(s=(P+k)^{2}\) can be defined in terms of the virtual photon four-momentum \(q\) as \cite{blumlein_2013},
\begin{equation} Q^{2} = -q^{2},\; y = \frac{P\cdot q }{P\cdot k}, \;  x = \frac{Q^{2}}{sy}  
\end{equation}

\section{SIDIS kinematic reconstruction}
In semi-inclusive DIS, observables are extracted in the nucleon center of mass frame, with the SIDIS cross-section a function of the inclusive DIS variables as well as (\(p_{h\perp}, z, \phi_h\)). The relevant transverse momentum is defined with respect to the virtual photon axis, and the single-hadron azimuthal angle \(\phi_h\) is defined between the lepton scattering plane and hadron production plane (figure 1). \(z\) is defined as 
\( z = \frac{p_{h}\cdot P}{q\cdot P}\)
The calculation of SIDIS kinematics therefore requires precise reconstruction of the four-momenta of the selected hadron and the exchanged virtual-photon.

\subsection{Electron method}
Extraction of SIDIS observables and multiplicities at the EIC presents a new challenge, as fully multi-dimensional SIDIS studies have so far only been carried out in lower energy fixed target experiments. In fixed target SIDIS studies, \(q\) has been determined using only the scattered electron, \(q = k - k'\). However, studies done for the EIC yellow report and EIC detector proposals have found that a significant contribution to uncertainty in SIDIS kinematics is poor reconstruction of the virtual photon four-momentum when using only the scattered electron. In particular, the electron method fails in such regions of kinematic phase space at the EIC such as at low y (\(y < 0.05\)), where the energy loss of the electron is small and not well resolved. This is a significant issue for the study of TMD effects at e-p colliders, as at low-\(Q^2\) and large-x spin-orbit correlations are expected to be most significant and higher twist effects are observable. Additionally, the low-y region will be critical for overlapping the phase space covered by the EIC and SIDIS studies carried out at other facilities such as Jefferson Lab.

\begin{figure}
    \centering
    \includegraphics[width=10cm]{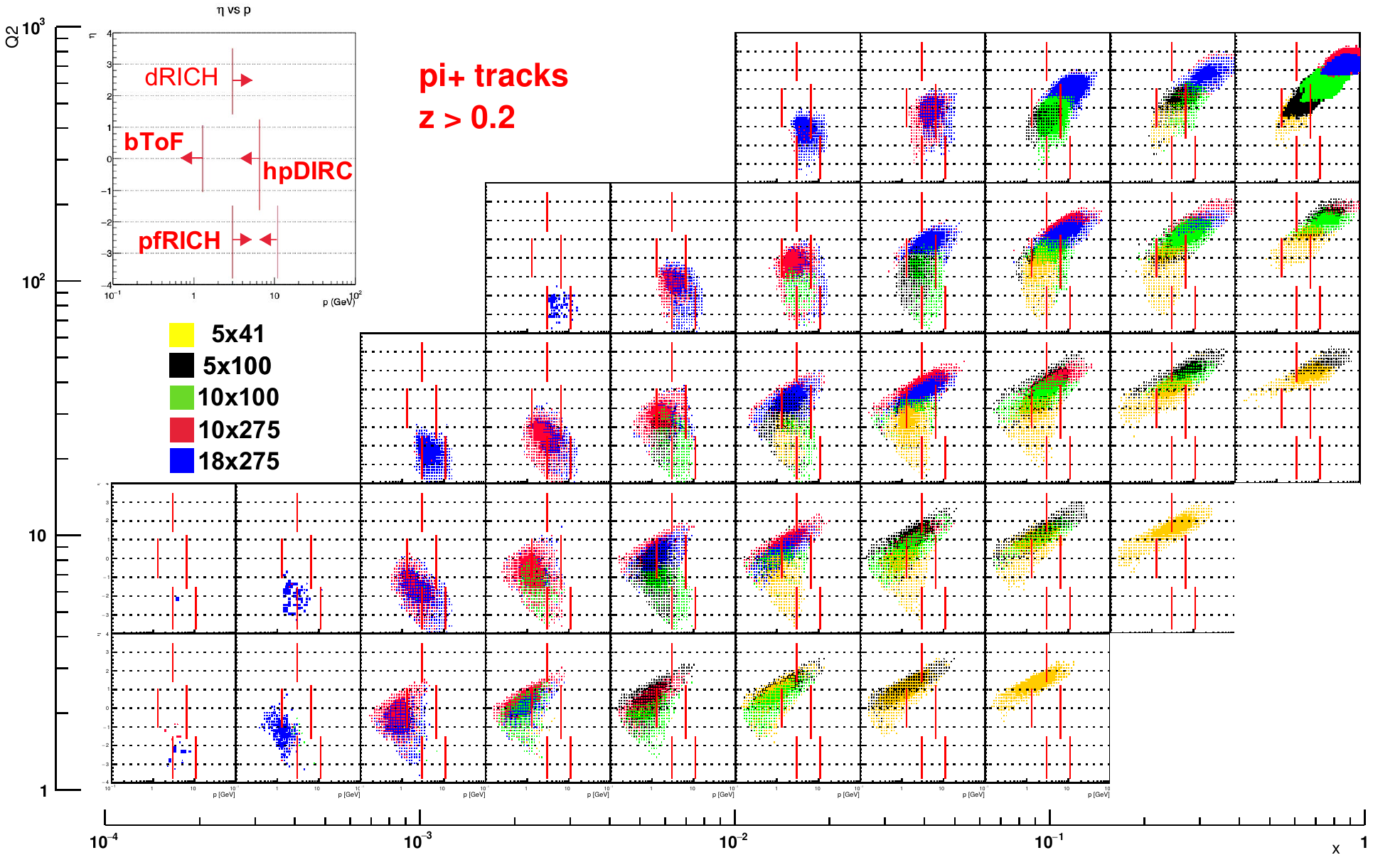}
    \caption{Momentum vs pseudorapidity in \(x-Q^{2}\) bins for positive pions with \(z > 0.2\) and requiring \(y > 0.05\), generated with Pythia-8. Different histogram colors represent different \(\sqrt{s}\) values. Red bands represent 3-\(\sigma\) PID coverage based on ATHENA proposal projections. }
    \label{fig:coverage}
\end{figure}

\subsection{Hadronic final state methods}
Fast simulation studies for the EIC yellow report~\cite{yellow_report} and ATHENA (A Totally Hermetic Electron Nucleon Apparatus) proposal~\cite{athena_proposal} have demonstrated that DIS reconstruction methods developed at past e-p colliders~\cite{blumlein_2013} can be used to improve the reconstruction of inclusive DIS variables at the EIC. The DIS reconstruction methods developed at HERA utilized combinations of measured quantities from the scattered electron and the hadronic final state (HFS). The use of the HFS allowed these additional methods, such as the double angle (DA) and \(\Sigma\)-methods ~\cite{blumlein_2013}, to improve inclusive DIS kinematic reconstruction for various regions of the HERA kinematic space, as well as to make the reconstruction robust with respect to QED radiative effects ~\cite{Bassler:1994uq,RevModPhys.71.1275}. For the studies planned at the EIC, methods utilizing the HFS must be extended to the reconstruction of SIDIS kinematics.

The authors of this contribution conducted first studies of SIDIS kinematic reconstruction for the EIC and demonstrated methods in which the hadronic final state can be used to improve the reconstruction of the virtual photon four momentum. This was carried out in the EIC yellow report and ATHENA proposal~\cite{yellow_report,athena_proposal} by first obtaining the transverse component of \(q\) from the recoil of the HFS transverse to the beamline through a sum of the momenta of HFS particles. Following the determination of this transverse recoil, the remaining two components of \(q\) can be constrained by the system of equations including \(q\) from the definitions of \(Q^2\) and \(y\),
\begin{equation} q_x = \sum_{i}^{N_{HFS}} p_{x,i}, q_y = \sum_{i}^{N_{HFS}} p_{y,i}\end{equation}
\begin{equation}
q_z, q_t \leftarrow
\begin{cases}
    Q^2 = -( q_x^2 + q_y^2 + q_z^2 - q_t^2 )\\
    y =  \frac{P_x q_x + P_y q_y + P_z q_z - P_t q_t}{P \cdot k}
\end{cases}
\end{equation}

In the EIC yellow report and ATHENA proposal \cite{yellow_report,athena_proposal}, this procedure was carried out using various inclusive DIS reconstruction methods developed at HERA\cite{blumlein_2013}, in fast simulations showing improvements over the electron method in some regions of the DIS kinematic space. As methods such as the Jacquet-Blondel (JB) method \cite{blumlein_2013} use only the hadronic final state information, this also allows for the determination of \(q\) from the HFS alone. Results using this approach are shown in the next section compared to ML and electron methods, with resolution using this method expected to improve with further developed full simulations based on fast simulation results.
\section{Machine learning kinematic reconstruction}
\subsection{Network architecture}
Multiple studies have been conducted demonstrating an improved resolution of inclusive DIS variables \(Q^2, y, x\) through deep learning approaches \cite{arratia_britzger_long_nachman_2022,diefenthaler_farhat_verbytskyi_xu_2021}, but these have not yet been extended to reconstruction of semi-inclusive DIS kinematics. In this study, we demonstrate that machine learning models which learn from the full HFS and scattered electron can be used to improve on current reconstruction methods to provide reliable reconstruction of the virtual photon axis across all of the DIS kinematic coverage at the EIC.

This approach to semi-inclusive DIS reconstruction is centered on the use of deep neural networks to better leverage the full hadronic final state at the level of reconstructed tracks. While previous applications of deep learning to inclusive DIS reconstruction directly regressed the kinematic variables of interest~\cite{arratia_britzger_long_nachman_2022,diefenthaler_farhat_verbytskyi_xu_2021}, this study aims to improve kinematics by directly regressing the virtual photon four-momentum in the lab frame.

Improvements to the HFS reconstruction are carried out through the use of Particle Flow Networks~\cite{Komiske:2018cqr}. Particle Flow Networks are an application of the deep sets neural network architecture, which learns a function of an unordered set of objects rather than from a fixed size input. The network consists of fully connected linear neural network layers which take as input the features of each particle individually, the outputs of which are summed over all particles to create a latent space representation of the event. The latent space variables and supplied global features of the event are then passed to another set of dense layers which produce the final output of the network ~\cite{Komiske:2018cqr}. Particle flow networks have seen particular success in tasks such as jet classification at the LHC. Particle flow networks implemented in Keras \cite{chollet2015keras} are included in the EnergyFlow python package. ~\cite{Komiske:2018cqr}
\begin{figure}
    \centering
    \includegraphics[width=8cm]{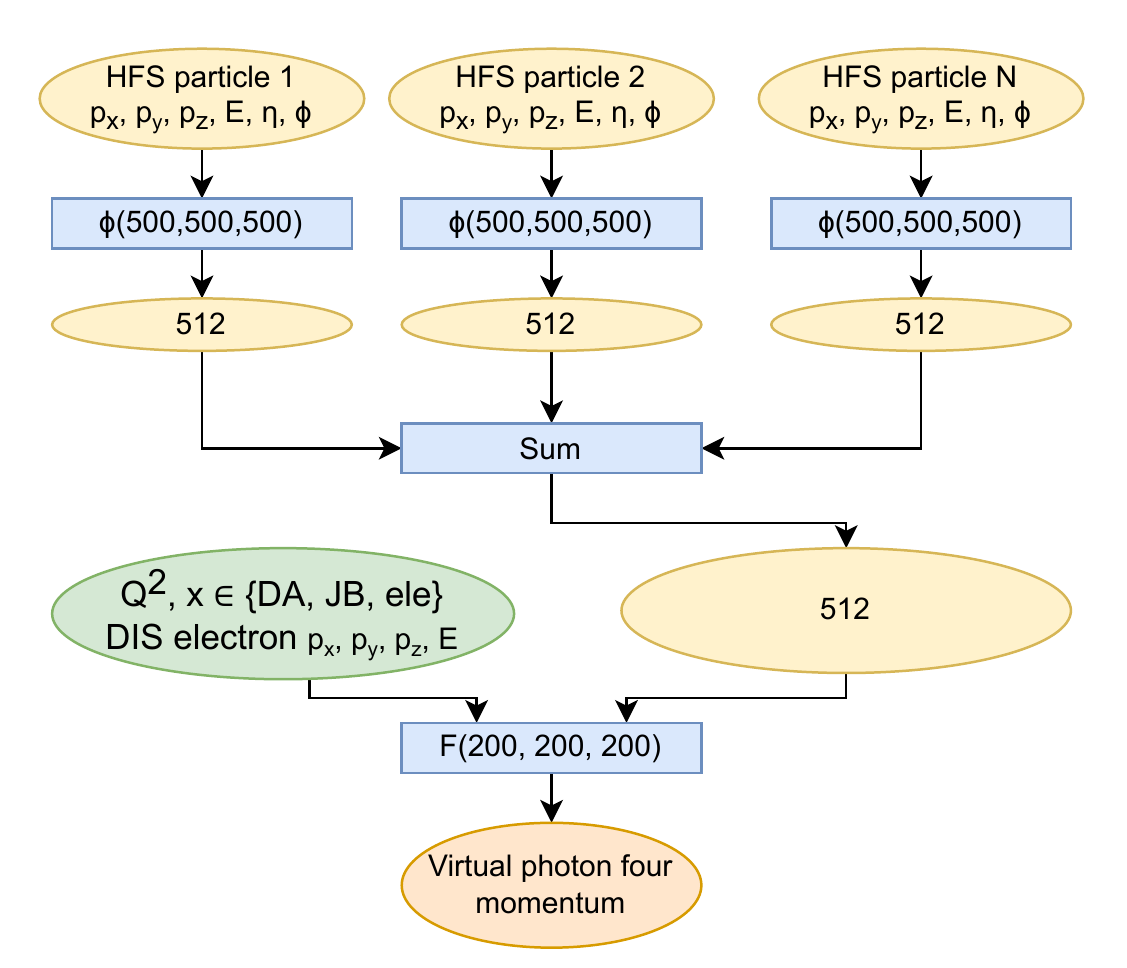}
    \caption{Network diagram of Particle Flow Networks \cite{Komiske:2018cqr} with global event features. Features of each HFS particle supplied individually to layers \(\Phi\), then summed over to form a latent space representation. Latent space features and global features of event (green), including reconstructed inclusive DIS variables and DIS electron momentum, supplied to layer \(F\) which produces final output. }
    \label{fig:pfn_diagram}
\end{figure}

\subsection{Variables and dataset}
\begin{figure*}
    \centering
    \includegraphics[width=8cm]{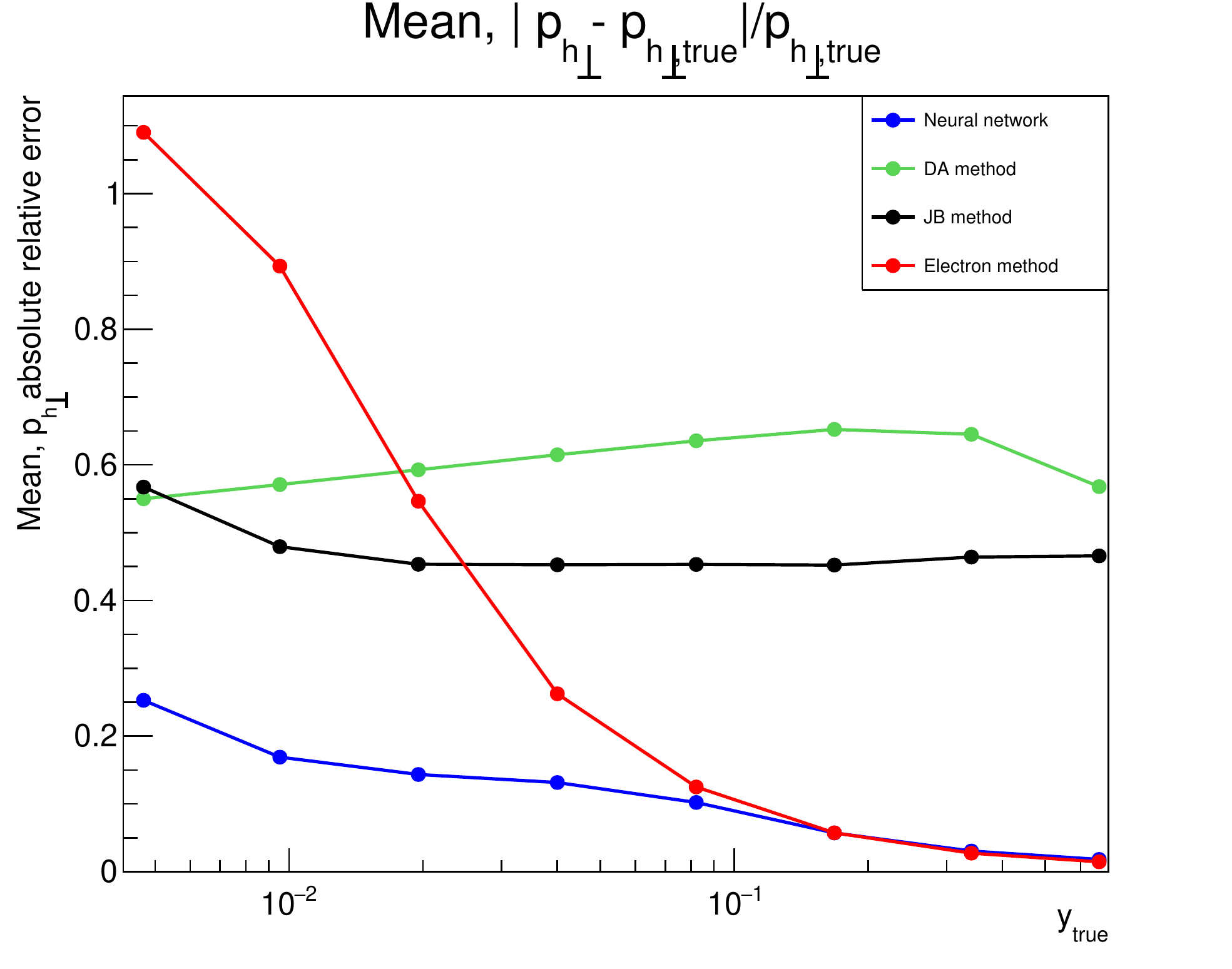}
    \includegraphics[width=8cm]{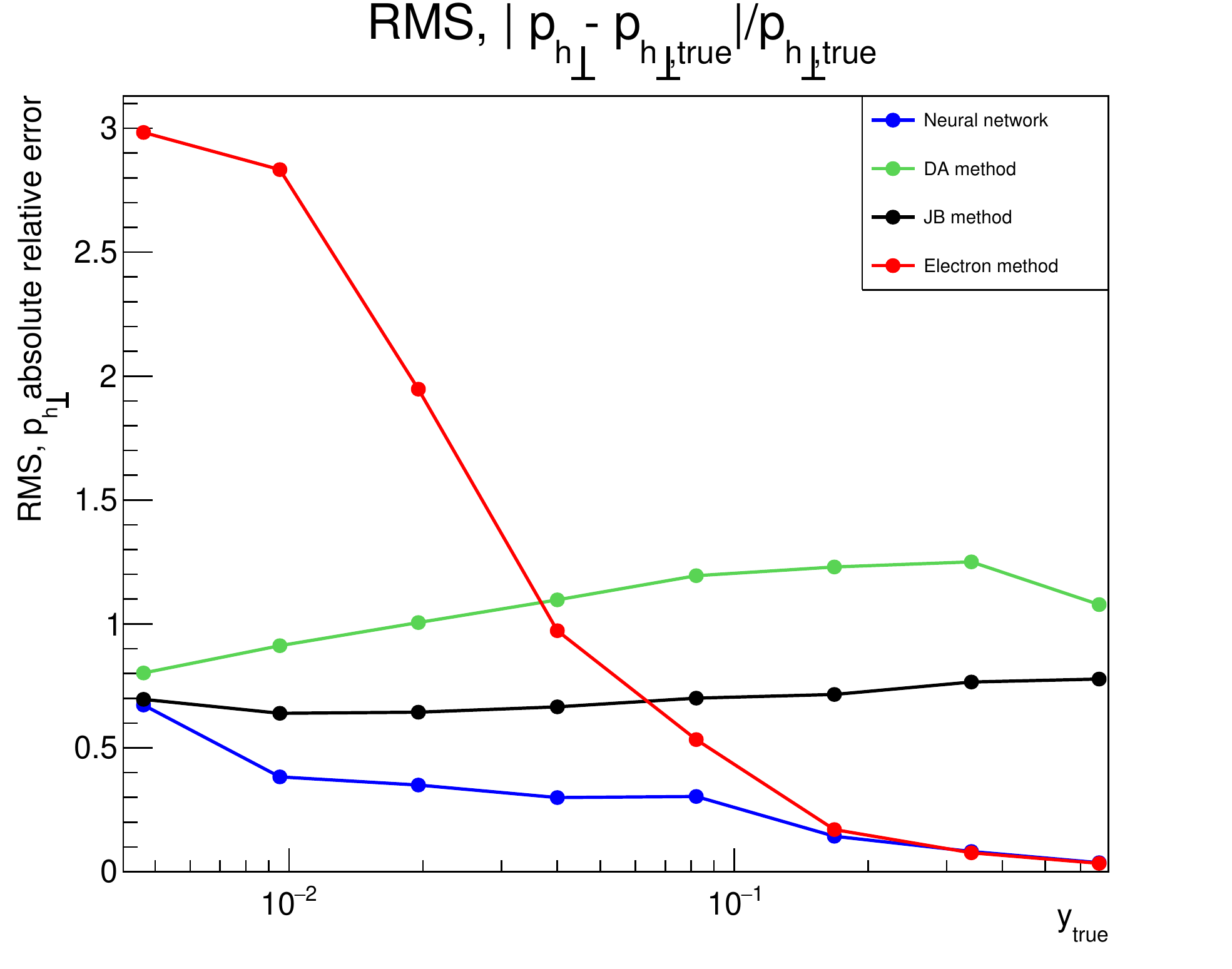}
    \caption{SIDIS \(p_{h\perp}\) resolution mean (left) and RMS (right) as a function of \(y_{true}\) for positive pions with \(z>0.2\), \(p_{h\perp} > 0.1 GeV\). HFS methods surpass electron method for very low \(y\), while PFN equals or outperforms electron method for all \(y\). }
    \label{fig:fullvsy_pt}
\end{figure*}
\begin{figure*}
    \centering
    \includegraphics[width=8cm]{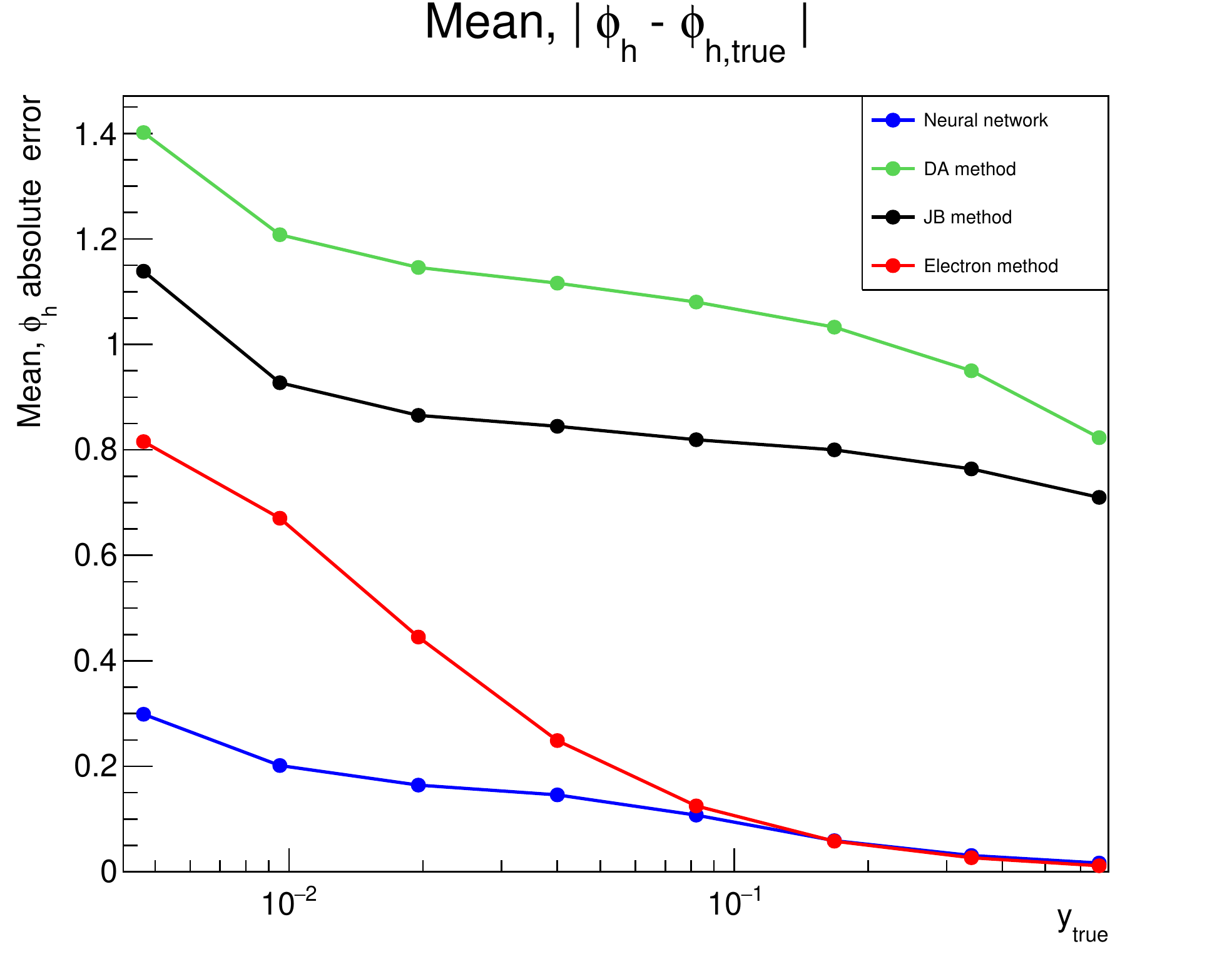}
    \includegraphics[width=8cm]{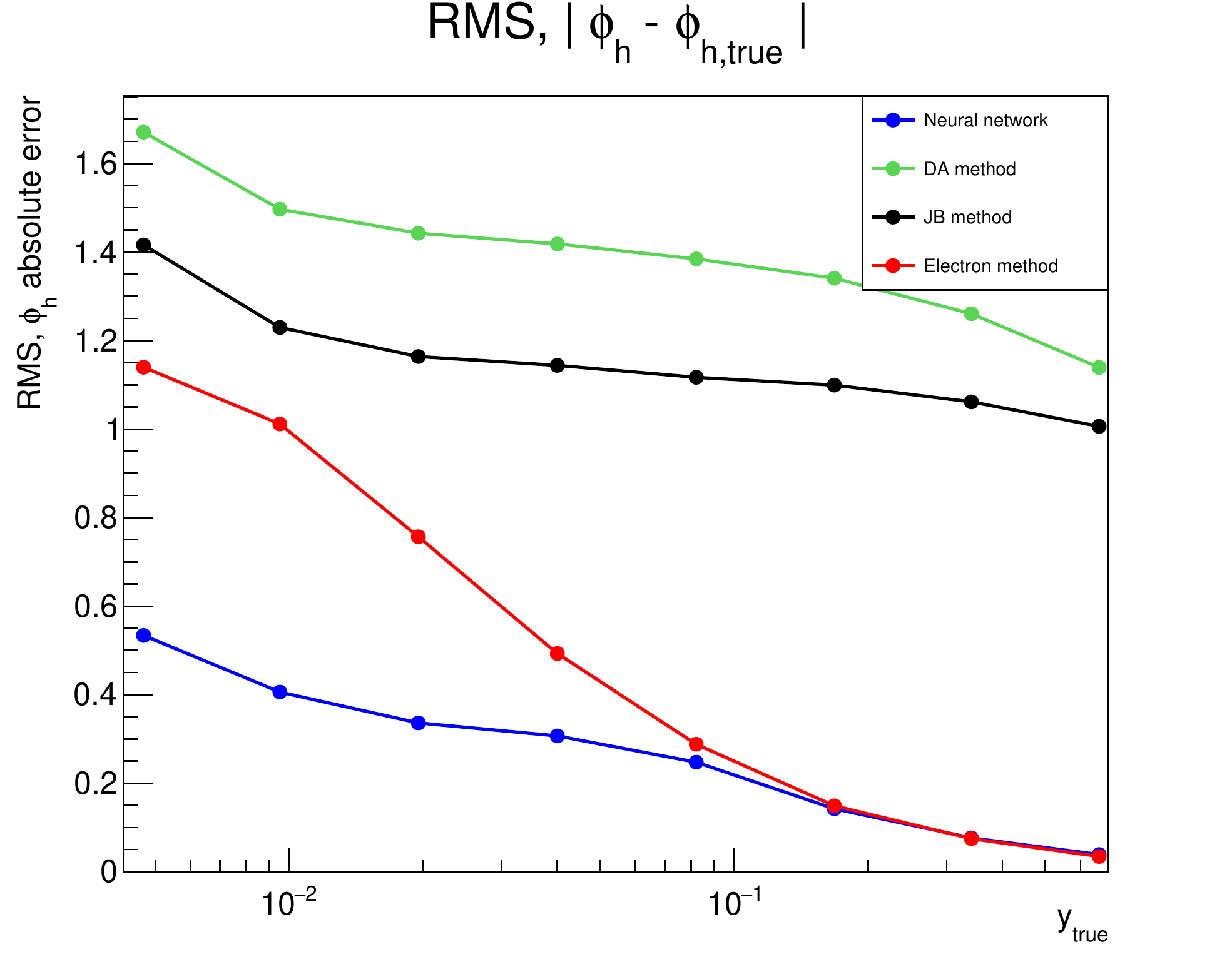}
    \caption{SIDIS \(\phi_{h}\) resolution mean (left) and RMS (right) as a function of \(y_{true}\) for positive pions with \(z>0.2\), \(p_{h\perp} > 0.1 GeV\). }
    \label{fig:fullvsy_phi}
\end{figure*}
The features of the hadronic final state reconstructed particles provided to the particle flow network include the four-momentum of each particle, as well as the lab frame azimuthal angle and pseudorapidity to provide direct information on angular acceptance in addition to momentum in each direction. 

The global features used for training include the four-momentum of the scattered electron and the DIS variables \(x\) and \(Q^2\) from the electron, DA, and JB methods. By supplying the full electron four-momentum following the single-particle layers \(\Phi\), the model is intended to learn corrections to the electron method based on the hadronic final state latent space variables. When a greater amount of fully simulated EIC simulated data is available, the DIS methods could also be replaced by the output of the deep learning models for inclusive DIS variables described previously.

The particle flow network was trained to predict the full four-momentum \(q\) in the lab frame. The particle flow network, implemented in Keras \cite{chollet2015keras} and available in the EnergyFlow python package, is used with per-particle dense layer units \(\phi = (500, 500, 500)\), \(l = 512\), and final dense layer units \(F = (200, 200, 200) \). Both the layers making up \(\phi\) and \(F\) employ a relu activation function, with the final output layer having linear activation.
\begin{figure*}
    \centering
    \includegraphics[width=8cm]{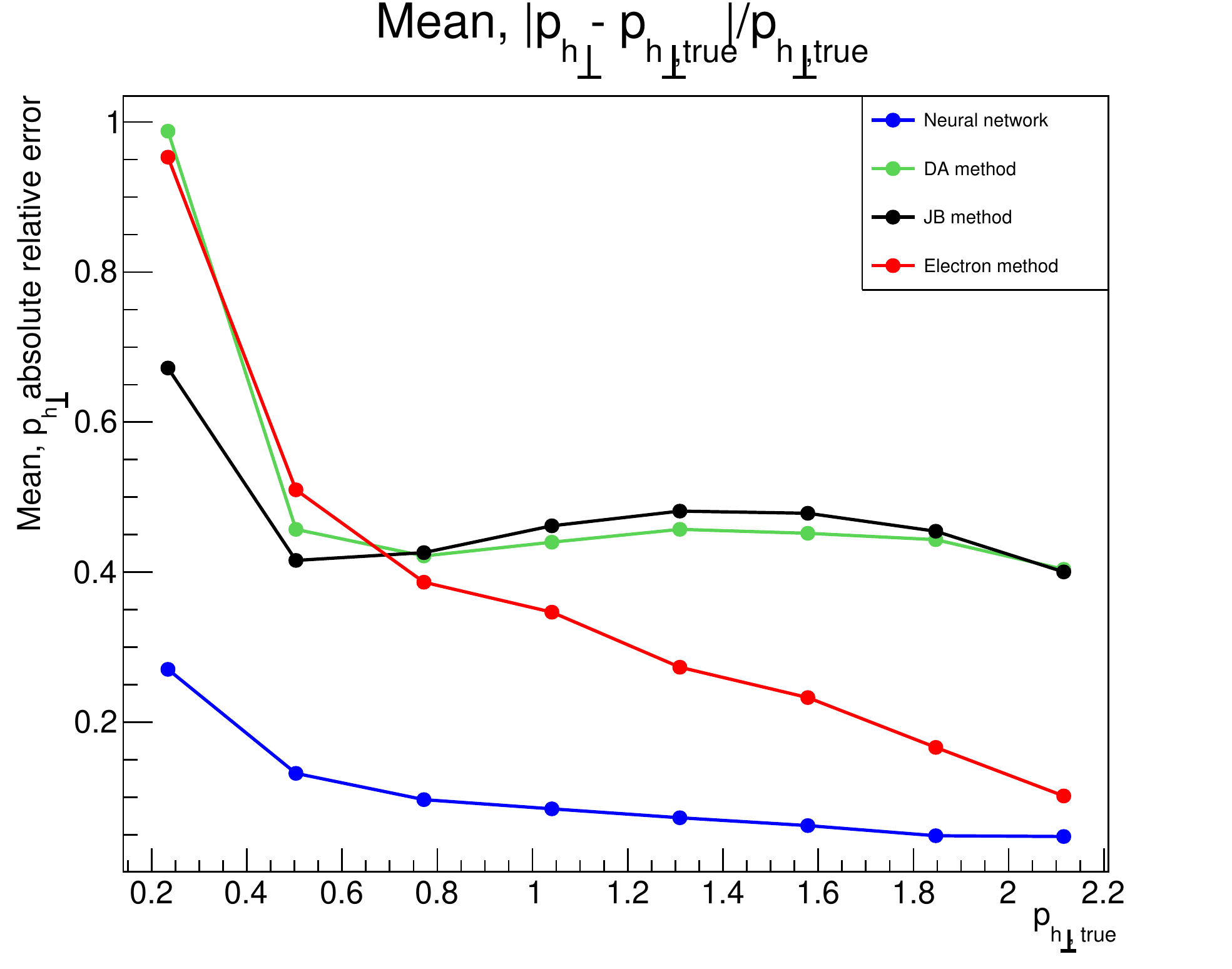}
    \includegraphics[width=8cm]{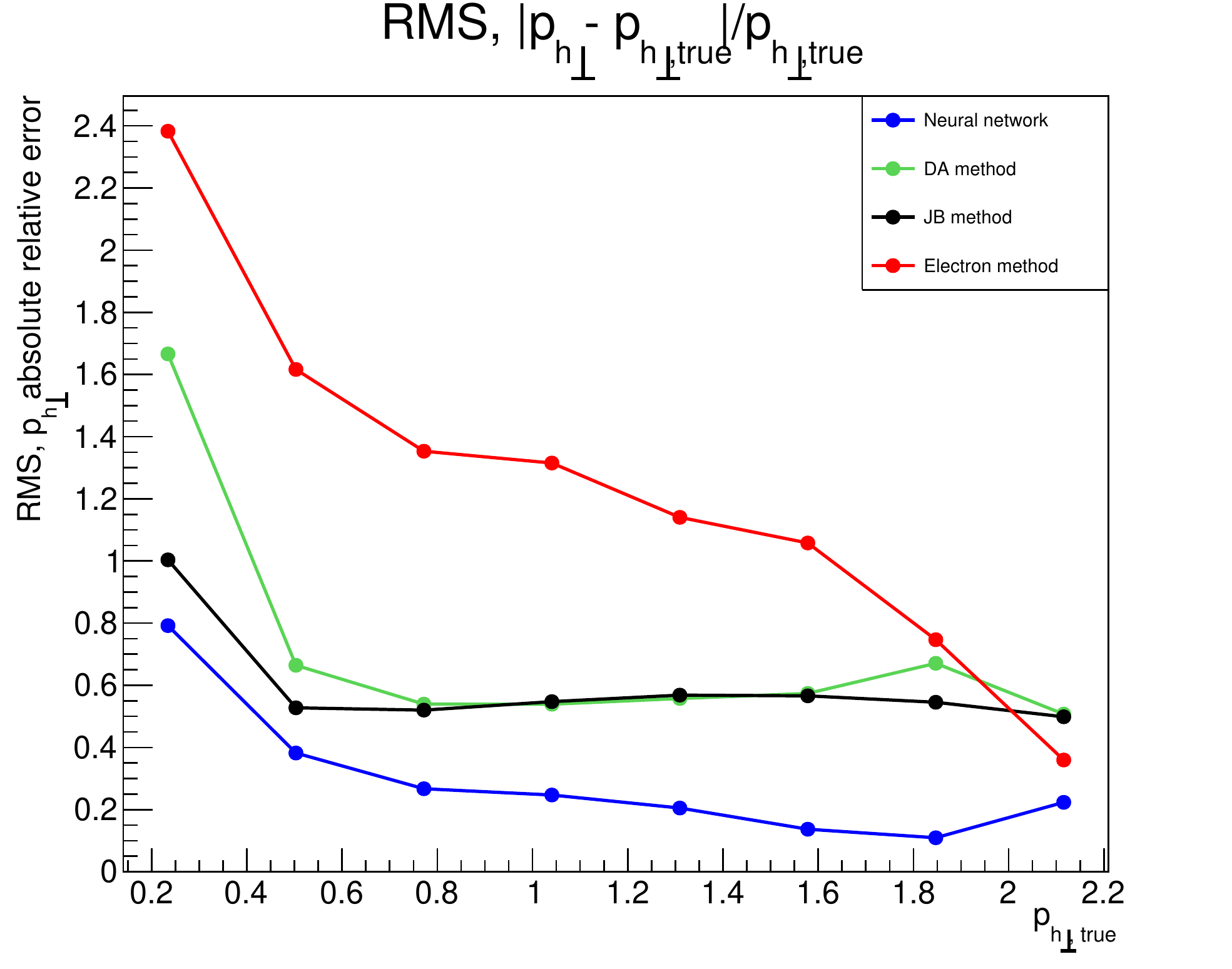}
    \caption{SIDIS \(p_{h\perp}\) resolution mean (left) and RMS (right) as a function of \(p_{h\perp,true}\) for positive pions with \(z>0.2\). }
    \label{fig:fullvspT_largey_pT}
\end{figure*}

\begin{figure*}
    \centering
    \includegraphics[width=8cm]{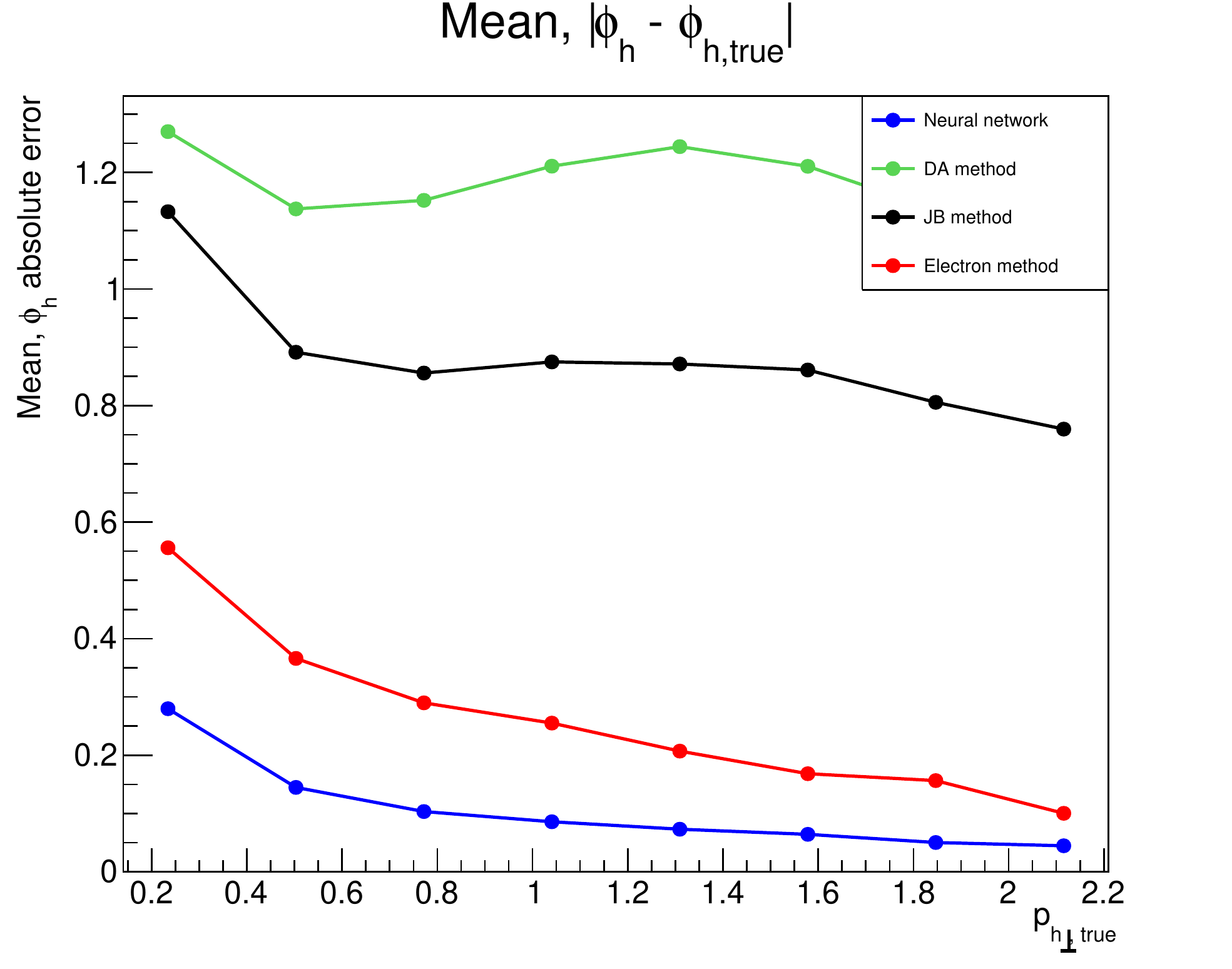}
    \includegraphics[width=8cm]{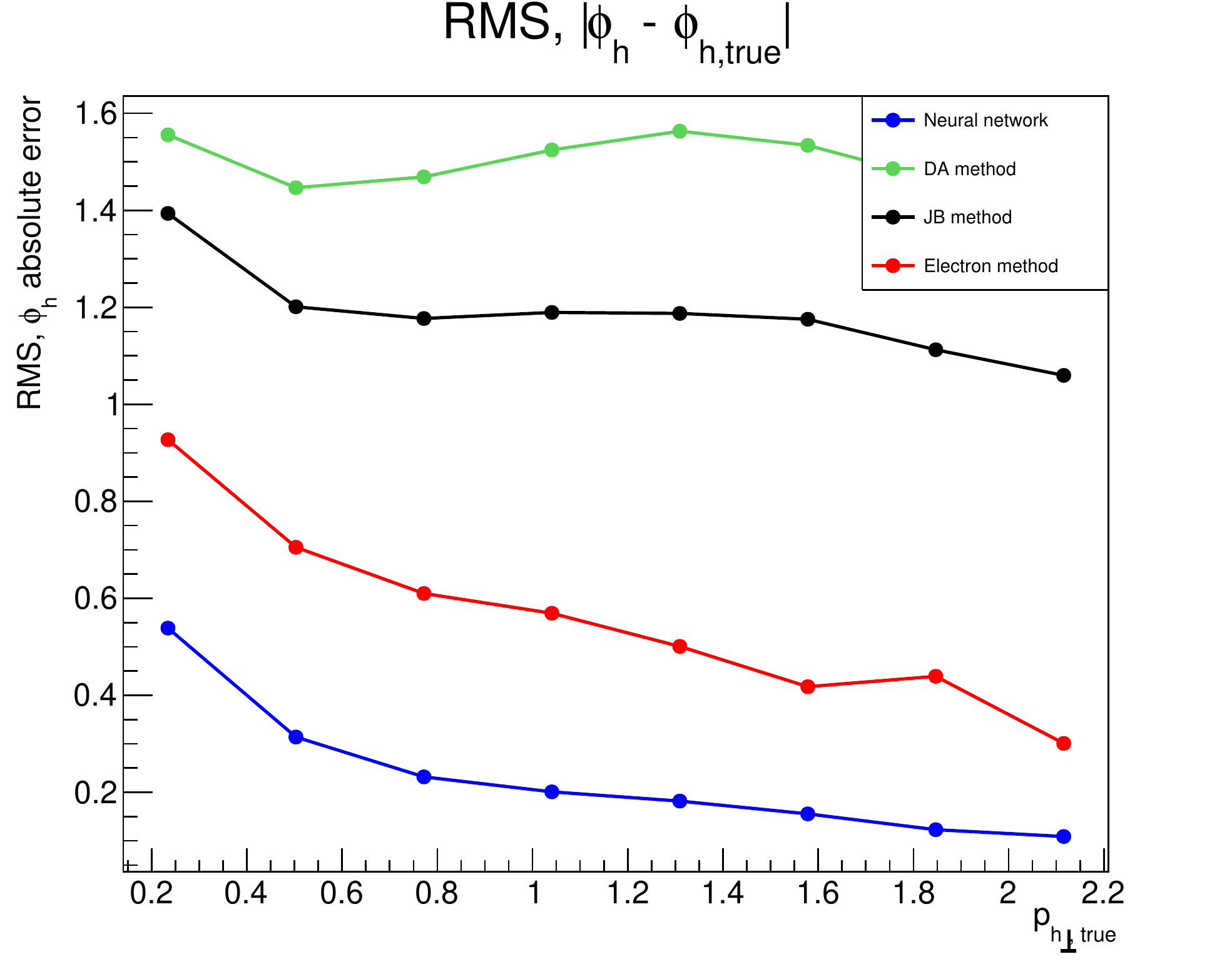}
    \caption{SIDIS \(\phi_{h}\) resolution mean (left) and RMS (right) as a function of \(p_{h\perp,true}\) for positive pions with \(z>0.2\). }
    \label{fig:fullvspT_largey_phi}
\end{figure*}

The dataset used for the training and testing of this model was the ATHENA full simulation developed for the ATHENA detector proposal for the first interaction region at the EIC. ATHENA was developed with the objective of meeting the resolution and physics goals laid out in the EIC yellow report. The ATHENA full simulation was implemented in DD4hep, Geant4, and Juggler \cite{frank_markus_2018_1464634,ALLISON2016186,jugglerGit}. At the time of the detector proposal, PID algorithms were not fully implemented, meaning PID information was not included in this model. Additionally, the scattered electron was taken as always correctly identified by matching the scattered electron with the MC truth information.

The simulated event sample used for model training and testing was a neutral current DIS sample generated using Pythia-8 \cite{pythia-manual-2022}, with additional beam smearing and crossing angle effects implemented. 3 million events with \(Q^2 > 1 \: \mathrm{GeV}^2\) and 2 million events with \(Q^2 > 10 \: \mathrm{GeV}^2\) were used for training with 1 million \(Q^2 > 1 \: \mathrm{GeV}^2\) set aside for model validation.

\section{Results}
As a function of \(y\) (figures 4 and 5), using the virtual photon four-momentum as predicted by the neural network model results in significantly improved reconstruction of \(p_{h\perp}\), \(\phi_h\) and \(z\) for low-y, when compared to both the electron method and methods utilizing information from the hadronic final state. The neural network reconstruction of \(q\) results in a distribution of the SIDIS variables which is both better centered around the true value, and with a significantly smaller RMS where the electron method begins to fail at low-y. At large-y, the neural network achieves performance only slightly surpassing that of the electron method, which is expected based on the projected energy and tracking resolution for the scattered electron with ATHENA. 

As a function of \(p_{h\perp,true}\), we also observe a significant improvement in kinematic reconstruction for both transverse momentum and for the semi-inclusive azimuthal angle. As the electron method begins to degrade for lower values of \(p_{h\perp}\), the neural network reconstruction of \(q\) results in stable performance to the lowest values of \(p_{h\perp}\) in the dataset.

\section{Summary}
The EIC will provide the first opportunity for semi-inclusive DIS measurements in an e-A collider context, giving access to new kinematic regions in which to precisely explore the 3-dimensional spin structure of nucleons. The development of reliable kinematic reconstruction methods will be critical to enabling precision extraction of SIDIS observables, especially at low-y. This can be achieved through the use of information from the hadronic final state alongside the scattered electron. As demonstrated in this contribution, machine learning, here using particle flow networks, can combine the information from the scattered electron and full HFS to provide reliable SIDIS kinematic reconstruction across the DIS variable space. Further steps in this work will include the consideration of QED radiative effects on SIDIS reconstruction, as well as possible extension to other neural network architectures exploiting correlations between particles. Additionally, this approach will continue to be studied and validated as more detailed full detector simulations are developed for the EIC. 

\section{Acknowledgements}
We thank Markus Diefenthaler for helpful discussions throughout the development of the described machine learning approach and for comments on this submission.
\clearpage
\bibliography{ref}

\end{document}